\documentclass[prl,floats,twocolumn,aps,superscriptaddress,showpacs]{revtex4}

\usepackage{amsmath}
\usepackage{graphicx}
\usepackage{color}

\begin{document}

\title{Fluctuations of internal energy flow in a vibrated granular gas.}

\date{\today}

\author{Andrea Puglisi}
\affiliation{Laboratoire de Physique Th\'eorique (CNRS UMR8627), B\^atiment
  210, Universit\'e Paris-Sud, 91405 Orsay cedex, France}

\author{Paolo Visco}
\affiliation{Laboratoire de Physique Th\'eorique (CNRS UMR8627), B\^atiment
  210, Universit\'e Paris-Sud, 91405 Orsay cedex, France}
\affiliation{Laboratoire de Physique Th\'eorique
et Mod\`eles Statistiques (CNRS
UMR 8626), B\^atiment 100, Universit\'e Paris-Sud, 91405 Orsay cedex, France}

\author{Alain Barrat}
\affiliation{Laboratoire de Physique Th\'eorique (CNRS UMR8627), B\^atiment
  210, Universit\'e Paris-Sud, 91405 Orsay cedex, France}

\author{Emmanuel Trizac}
\affiliation{Laboratoire de Physique Th\'eorique
et Mod\`eles Statistiques (CNRS
UMR 8626), B\^atiment 100, Universit\'e Paris-Sud, 91405 Orsay cedex, France}

\author{Fr\'ed\'eric van Wijland}
\affiliation{Laboratoire de Physique Th\'eorique (CNRS UMR8627), B\^atiment
  210, Universit\'e Paris-Sud, 91405 Orsay cedex, France}
\affiliation{P\^ole Mati\`ere et Syst\`emes Complexes 
(CNRS UMR 7057), Universit\'e
Denis Diderot (Paris VII), 2 place Jussieu, 75251 Paris cedex 05, France}

\begin{abstract}
  The non-equilibrium fluctuations of power flux in a fluidized
  granular media have been recently measured in an experiment [Phys.  Rev.
  Lett.  92, 164301, 2004], which was announced to be a verification of the
  Fluctuation Relation (FR) by Gallavotti and Cohen.
  An effective temperature was also identified and proposed to be
  a useful probe for such non equilibrium systems.
  We explain these
  results in terms of a two temperature Poisson process.
  Within this model, supported by independent Molecular Dynamics simulations,
  power flux fluctuations do not satisfy the FR and
  the nature of the effective temperature is clarified. 
  In the pursue of a hypothetical global quantity 
  fulfilling the FR, this points to the need of considering other
  candidates than the power flux.
\end{abstract}

\pacs{02.50.Ey, 05.20.Dd, 81.05.Rm}
\maketitle

%%%%%%%%%%%%%%%%%%%%%%%%%%%%%%%%%%%%%%%%%%%%%%%%%%%%%%%%%%%%%%%%%%%%%%%%%%%

Granular gases, i.e. gases of macroscopic grains losing part of their kinetic
energy during collisions, have been the subject of ardent investigation in the
last decade~\cite{intro1}. They display, in experiments as well as in
simulations, a rich and intriguing phenomenology: non-Gaussian statistics,
breakdown of energy equipartition, spontaneous symmetry breaking (clustering,
shear, convection, surface waves, shocks etc) are the most striking features
of this surprising state of matter~(see e.g.~\cite{barrat} and references
therein). In addition, a driven granular gas is an ideal testing ground for
non-equilibrium statistical mechanics of dissipative stationary states. An
inelastic gas can be kept in a state of constant average total kinetic energy
thanks to an external driving mechanism, e.g. by rapidly vibrating its
container. In such a system, energy is flowing from the external
source/thermostat into the gas and then from the gas into an irreversibly
draining sink, represented by inelastic collisions.  As a consequence, the
fluctuations of global or microscopic physical observables, such as the total
energy or the velocity of particles or all internal currents, do not behave as
expected in equilibrium statistical mechanics. It is therefore tempting to
give an interpretation of the observed fluctuations in terms of the few known
theoretical results in non-equilibrium statistical physics. In a recent
experiment on vibrated granular gases~\cite{feitosa} it has been argued that
the statistics of the power flux fulfilled the Fluctuation Relation (FR) by
Gallavotti and Cohen~\cite{gc,stochastic}, even if 
--as emphasized in \cite{feitosa}-- a) the FR holds under
conditions that are not met in the situation under study (in particular,
microscopic reversibility is required) and b) there is no proven relation
between the measured power flux and the entropy production entering the
original FR. 

We show here that these experimental results can be explained in terms of a
simple Poisson process. We argue that the measured ``power flux'' $Q$ is well
reproduced by a sum of random and independent energy amounts characterized by
two different typical temperatures. Such a quantity does not verify the FR. We
nevertheless demonstrate that a straight line with slope $\beta_{eff}$ can be
observed when plotting $\log[f_Q(Q)/f_Q(-Q)]$ vs. $Q$, ($f_Q$ being the
probability density distribution of $Q$), in the range which is accessible in
experiments and simulations.  In~\cite{feitosa} the predictability of
$\beta_{eff}$ remained an open problem and $1/\beta_{eff}$ was interpreted as
an effective temperature. Within the two temperature model, the value of
$\beta_{eff}$ is obtained and it further appears that interpreting
this quantity as an inverse effective temperature is problematic.
We validate our analysis by a direct confrontation against 
Molecular Dynamics (MD) simulations, obtaining very good agreement.

%%%%%%%%%%%%%%%%%%%%%%%%%%%%%%%%%%%%%%%%%%%%%%%%%%%%%%%%%%%%%%%%%%%%%%%%%%%%
\begin{figure}[htbp]
\begin{flushleft}
\makebox[4.2cm][l]{\includegraphics[height=4.2cm]{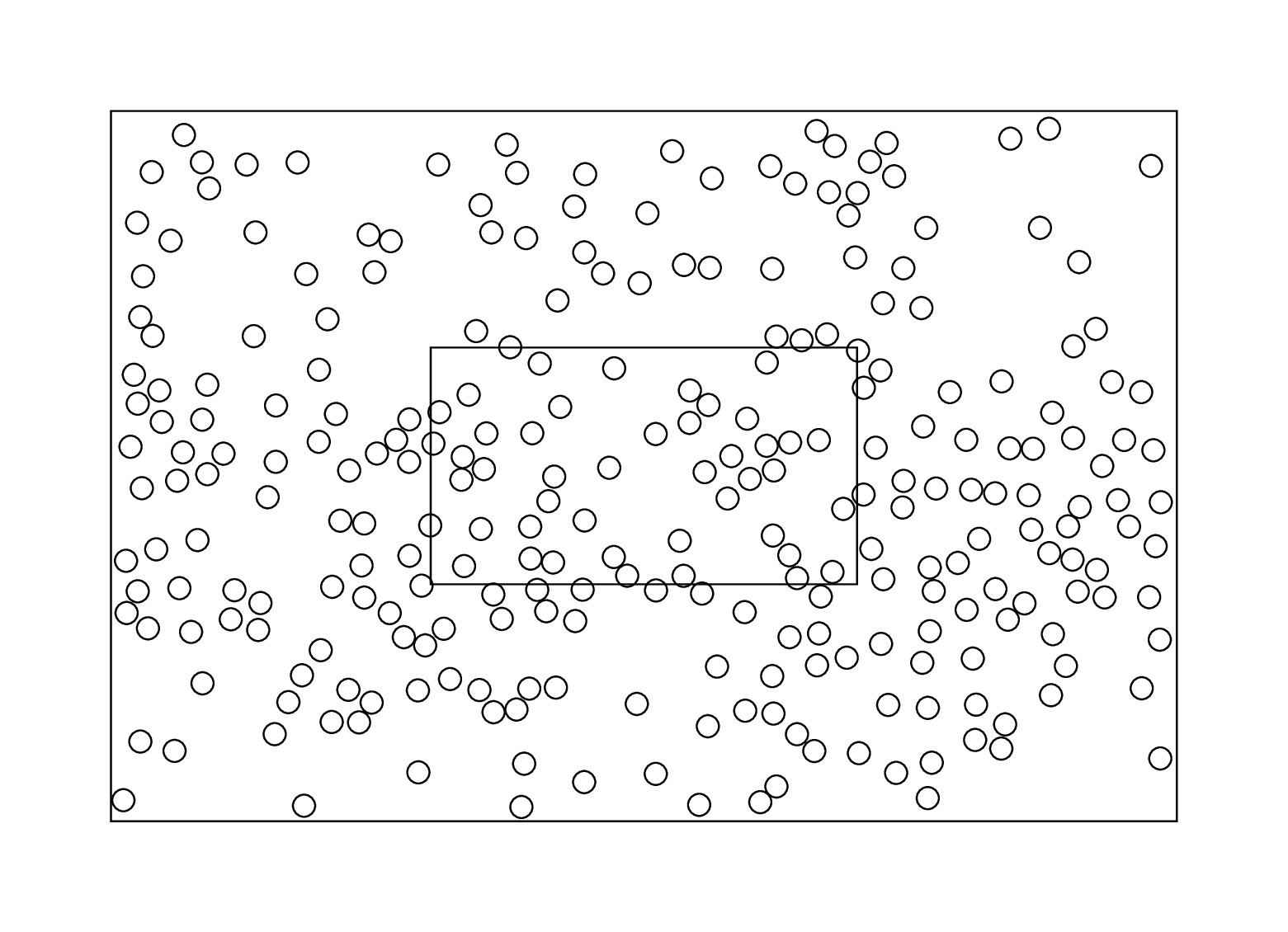}}
\makebox[4.3cm][r]{\includegraphics[height=4.2cm]{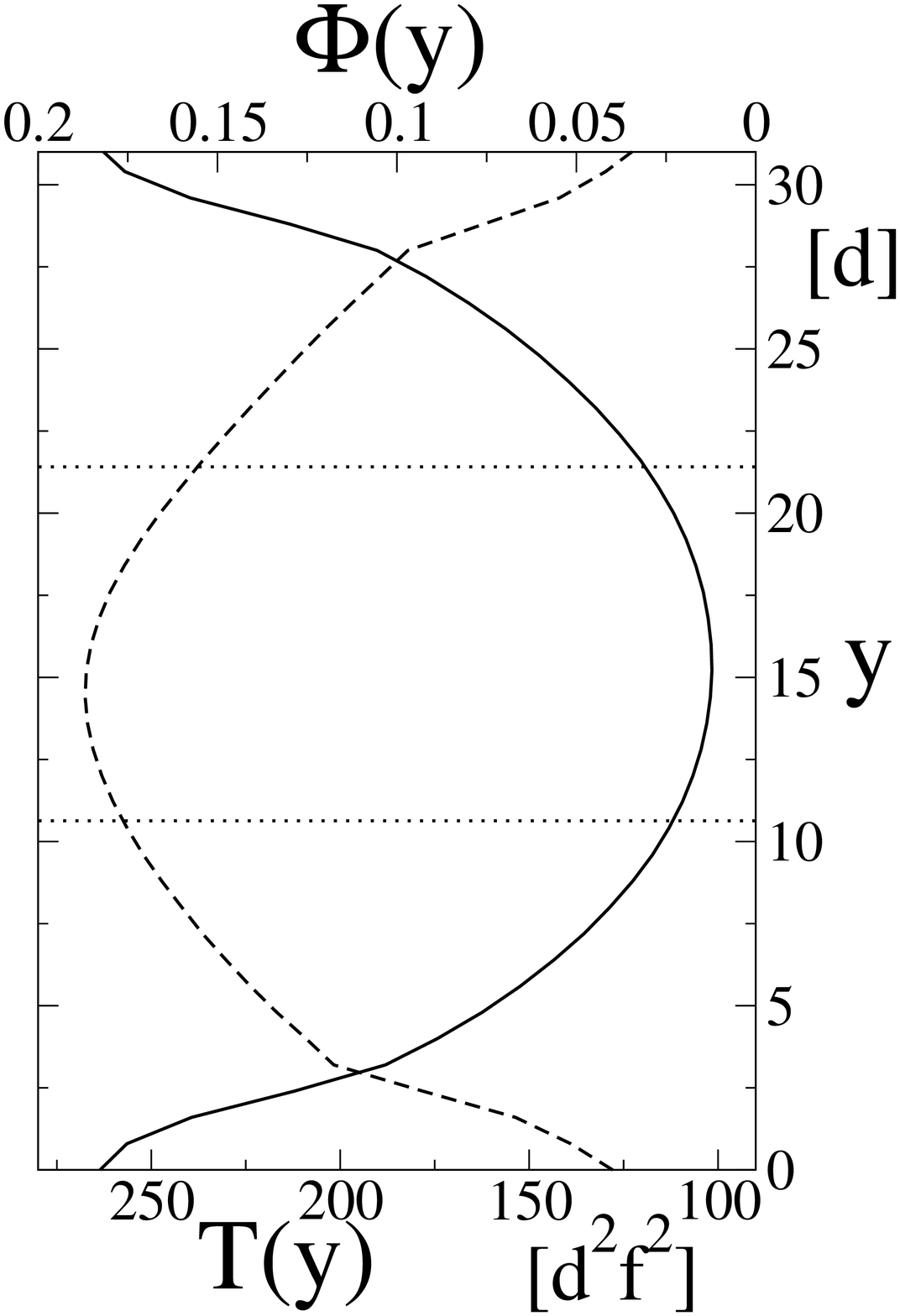}}
\end{flushleft}
\caption{Left: Snapshot of the system considered for MD simulations,
  with the inner region marked by the solid rectangle. Right:
  Corresponding vertical profiles of density ($\Phi(y)$, dashed line)
  and temperature ($T(y)$, solid line). The dotted lines mark the
  bottom and top boundaries of the inner region. Here $N=270$ and
  $\alpha=0.9$. The mean free path is $\sim 5.7d$.
 \label{fig:MD}}
\end{figure}

The event driven MD simulations have been performed for a system of
$N$ inelastic hard disks with restitution coefficient $\alpha$,
diameter $d$ and mass $m=1$.  The vertical $2D$ box of width $L_x=48d$
and height $L_y=32d$ is shaken by a sinusoidal vibration with
frequency $f$ (period $\tau_{box}=1/f$) and amplitude $2.6d$. In a
collision two particles lose a fraction $1-\alpha^2$ of their relative
kinetic energy while the total momentum is conserved.  Collisions with
the elastic walls inject energy and allow the system to reach a
stationary state.  We have checked that possible inelastic collisions
with the walls hardly affect the results. Gravity -- set to
$g=-1.7df^2$ in order to be consistent with the experiment-- has a
negligible influence on the measured quantities.  We have varied the
restitution coefficient from $0.8$ up to $0.99$ (glass beads yield on
average $\alpha \approx 0.9$) and the total area coverage from $0.138$
(i.e.  $N=270$) up to $0.32$ ($N=620$). In Fig.~\ref{fig:MD}-left a
snapshot of the system is shown.  During the simulations all the main
physical observables are statistically stationary.  The local area
coverage field $\Phi(x,y)$ and the granular temperature field $T(x,y)$
(defined in $2D$ as the average kinetic energy per particle) are
almost uniform in the horizontal direction, apart from small layers
near the side walls.  In Fig.~\ref{fig:MD}-right the profiles
$\Phi(y)=(1/L_x)\int dx \Phi(x,y)$ and $T(y)=(1/L_x)\int dx T(x,y)$
are shown to be symmetric with respect to the bottom and the top of
the box.  Following the experimental procedure, we have focused our
attention on a ``window'' in the center of the box, fixed in the
laboratory frame, of width $2L_x/5$ and height $L_y/3$, marked in
Fig.~\ref{fig:MD}-left. Apart from the negligible change of potential
energy due to gravity, the total kinetic energy of the particles
inside the window, changes during a time $\tau$ because of two
contributions: $\Delta K_\tau= Q_\tau - W_\tau$ where $Q_\tau$ is the
kinetic energy transported by particles through the boundary of the
window (summed when going-in and subtracted when going-out) and
$W_\tau$ is the kinetic energy dissipated in inelastic collisions
during time $\tau$.  For several values of $\tau$ we have measured, as
in the experiments, $Q_\tau$ which is related to the kinetic
contribution to the heat flux ({\color{blue} we checked that inclusion of the
collisional contribution, even if non small~\cite{HMZ}, does not change
the picture}).  With $N=270$ and $\alpha=0.9$ the characteristic times
are the mean free time $\tau_{col} \approx 0.47\tau_{box}$, the
diffusion time across the window $\tau_{diff}=0.82\tau_{box}$ and the
mean time between two subsequent crossings of particles (from outside
to inside) $\tau_{cross}\approx 0.039\tau_{box}$.

\begin{figure}[htbp]
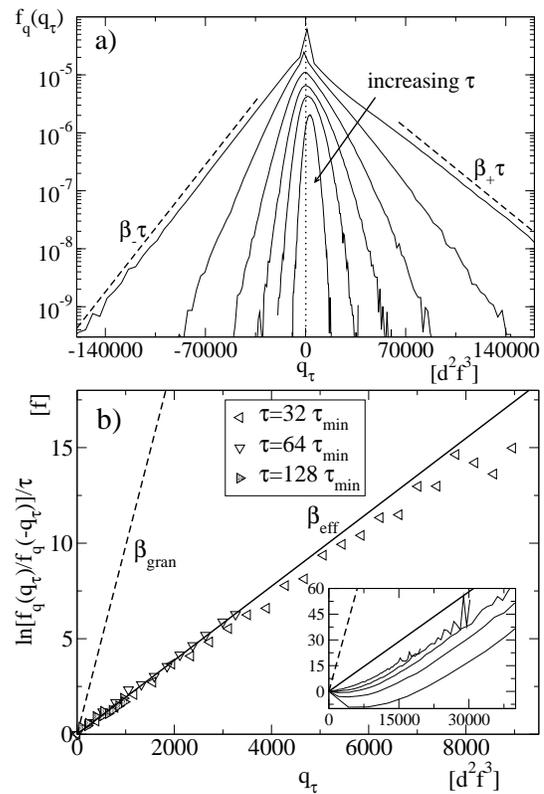

\includegraphics[width=7cm,clip=true]{pdf.eps}\\
\includegraphics[width=7cm,clip=true]{gc.eps}
\caption{{\bf a)} pdfs of injected power $f_q(q_\tau)$ from MD
  simulations for different values of $\tau=(1,2,4,8,16,32)\times
  \tau_{min}$ with $\tau_{min}=0.015 \tau_{box}$. Here $N=270$ and
  $\alpha=0.9$. The distributions are shifted vertically for
  clarity.  The dashed lines put in evidence the exponential tails of
  the pdf at $\tau=\tau_{min}$.  {\bf b) } plot of
  $(1/\tau)\log[f_q(q_\tau)/f_q(-q_\tau)]$ vs. $q_\tau$ from MD
  simulations (same parameters as above) at large values of $\tau$. The
  solid curve is a linear fit (with slope $\beta_{eff}$) of the data
  at $\tau=128 \tau_{min}$. The dashed line has a slope
  $\beta_{gran}=1/T_{gran}$. In the inset the same graph is shown for
  small values of $\tau=(1,2,4,8) \times \tau_{min}$ (from bottom to
  top).\label{fig:pdf}}
\end{figure}

We define the injected power as $q_\tau=Q_\tau/\tau$ and two relevant
probability density functions (pdfs): $f_Q(Q_\tau)$ and $f_q(q_\tau)$.
Fig.~\ref{fig:pdf}a) shows $f_q(q_\tau)$ for different values of $\tau$. A
direct comparison with Fig.~3 of Ref.~\cite{feitosa} suggests a fair agreement
between simulations of inelastic hard disks and the experiment. The pdfs are
strongly non-Gaussian and asymmetric, becoming narrower as $\tau$ is
increased. At small $\tau$ a strong peak in $q_\tau=0$ is visible.  More
interestingly, $f_q(q_\tau)$ at small values of $\tau$ has two different
exponential tails, i.e.  $f_q(q_\tau) \sim \exp(\mp\beta_\pm\tau q_\tau)$ when
$q_\tau \to \pm\infty$ with $\beta_->\beta_+$.  The peak and the exponential
tails at small $\tau$ are observed also in the experiment (see Fig.~3
of~\cite{feitosa}) and in similar simulations~\cite{aumaitre}. In
Fig.~\ref{fig:pdf}b) we display $\log[f_q(q_\tau)/f_q(-q_\tau)]/\tau$ vs.
$q_\tau$, which is equivalent to the graph of $\pi(q_\tau)-\pi(-q_\tau)$ vs.
$q_\tau$ where $\pi(q_\tau)=\log[f_Q(\tau q_\tau)]/\tau$. When $\tau \to
\infty$, $\pi(q_\tau) \to \Pi(q)$, i.e. the large deviation function
associated to $f_Q(Q_\tau)$. Under a number of hypothesis,
the Gallavotti-Cohen Fluctuation Theorem \cite{gc} states
that for the entropy production $\sigma$ (defined in a dynamical system as the
phase space contraction rate) $\Pi(\sigma)-\Pi(-\sigma) = \sigma$.  From
Fig.~\ref{fig:pdf} it appears that at large values of $\tau$,
$\pi(q_\tau)-\pi(-q_\tau)$ is linear with a $\tau$-independent slope $\beta_{eff}
\neq 1$. We have measured $\beta_{eff}$ with various choices of the restitution
coefficient $\alpha$ and of the covered area fraction finding similar results.
Ref. \cite{feitosa} reports $\beta_{eff} T_{gran} \sim 0.25$
where $T_{gran}$ is the mean granular temperature in the observation window.
Similar values are measured in our MD simulations. At area fraction $13.8\%$
and $\alpha=0.9$ we have
$\beta_{eff} T_{gran} \approx 0.23
$. At fixed $\alpha$
and increasing area fraction, $\beta_{eff} T_{gran}$ slightly increases, as in the
experiment.
As $\alpha \to 1$ the slope $\beta_{eff}$ decreases. At $\alpha=1$ (without gravity
and external driving) the distribution of $Q_\tau$ is symmetrical and
$\beta_{eff}=0$, indicating that $1/\beta_{eff}$ is not a physically relevant
temperature concept. Interestingly, it appears that $\beta_{eff}$
is a non hydrodynamic quantity: different systems may show the {\em same}
density and temperature profiles, with very different values of 
$\beta_{eff}$ \cite{puglisi2}.

We now adopt a coarse-grained description of the experiment which is
able to entirely capture the observed phenomenology. The measured flow of
energy is given by
\begin{equation}
Q_\tau=\frac{1}{2}\left(\sum_{i=1}^{n_+} v_{i+}^2-
\sum_{i=1}^{n_-} v_{i-}^2\right),
\end{equation}
where $n_-$ ($n_+$) is the number of particles leaving (entering) the window
during the time $\tau$, and $v_{i-}^2$ ($v_{i+}^2$) are the squared moduli of
their velocities. In order to analyze the statistics of $Q_{\tau}$ we take
$n_-$ and $n_+$ as Poisson-distributed random variables with average
$\omega\tau$, where $\omega$ corresponds to the inverse of the crossing time
$\tau_{cross}$. In doing so we neglect correlations among particles entering
or leaving successively the central region. A key point, supported by direct
observation in the numerical experiment, lies in the assumption that the
velocities ${\bf v}_{i+}$ and ${\bf v}_{i-}$ come from populations with
different temperatures $T_+$ and $T_-$ respectively. Indeed, compared with the
population entering the central region, those particles that leave it have
suffered on average more inelastic collisions, so that $T_-<T_+$.  Finally we
assume Gaussian velocity pdfs~\cite{nongauss}.  
%In the following we refer to this interpretation as to a two-temperature model.
Within such a framework,
the distribution $f_Q(Q_\tau)$ of $Q_\tau$ can be studied analytically. Here
it is enough to recall that $\frac{1}{2}\sum_{i=1}^n v_{i}^2$, in $D$
dimensions, if each component of ${\bf v}_i$ is independently
Gaussian-distributed with zero mean and variance $T$, is a stochastic variable
$x$ with a distribution $\chi_{n,T}(x)=f_{1/T,Dn/2}(x)$, where
$f_{\alpha,\nu}(x)$ is the Gamma distribution, and whose generating function
reads $\tilde{\chi}_{n,T}(z)=\left(1-Tz\right)^{-Dn/2}$~\cite{feller}.  It is
then straightforward to obtain the generating function of $Q_\tau$ in the form
$\tilde{f}_Q(z)=\exp [\tau\mu(z)]$ with
\begin{equation}
\mu(z)=\omega\left(-2+(1-T_+z)^{-D/2}+(1+T_-z)^{-D/2}\right).
\end{equation}
We observe that $\tilde{f}_Q(z)$ has two poles in $z=\pm1/T_\pm$ and
two branch cuts on the real axis for $z>1/T_+$ and $z<-1/T_-$. From $\mu(z)$
we immediately obtain the cumulants of $f_Q(Q_\tau)$ through the formula
$\langle Q^n \rangle_c=\tau\frac{d^n}{dz^n}\mu(0)$.

For $\tau \to \infty$ the large deviation theory states that $f_Q(Q_\tau) \sim
\exp(\tau \Pi(Q_\tau/\tau))$ and $\Pi(q)$ can be obtained from $\mu(z)$ through
a Legendre transform, i.e. $\Pi(q)=\smash{\underset{z}{max}}(\mu(z)-qz)$. 
The study of
the  singularities of $\mu(z)$ reveals the behavior of the large deviation
function $\Pi(q)$ for $q \to \pm\infty$. It can be seen that
\begin{align} \label{tails}
\Pi(q) &\sim -\frac{q}{T_+} \;\;(q \to \infty), \;\;\;
\Pi(q) &\sim \frac{q}{T_-} \;\;(q \to -\infty).
\end{align}
We emphasize however that it is almost impossible to appreciate these tails in
simulations and in experiments, since the statistics for large values
of $q$ and $\tau$ is very poor.
%We shall see that the same behavior can be
%observed at small values of $\tau$.

A Gallavotti-Cohen-type relation~\cite{gc,stochastic},
e.g. $\Pi(q)-\Pi(-q)=\beta q$ for any $q$ and an arbitrary value of
$\beta$ would imply $\mu(z)=\mu(\beta -z)$. One can see
that such a $\beta$ does not exist, i.e. the fluctuations of $Q_\tau$
do not satisfy a Gallavotti-Cohen-like relation.  The observed
linearity of the graph
$\log[f_q(q_\tau)/f_q(-q_\tau)]/\tau=\pi(q_\tau)-\pi(-q_\tau)$
vs. $q_\tau$ can be explained by the following
observation~\cite{aumaitre}: at large values of $\tau$ it is extremely
difficult, in simulations as well as in experiments, to reach large
values of $q$, while for small $q$, $\pi(q)-\pi(-q) \approx
2\pi'(0)q+o(q^3)$, i.e. a straight line with a slope
$\beta_{eff}=2\pi'(0)$ is likely to be observed. It has been already
shown~\cite{farago} that in dissipative systems deviations from the FR
can be hidden by insufficient statistics at high values of $q$.  The
knowledge of $\mu(z)$ is useful to predict this slope.  
At large $\tau$, $\pi'(0)\approx\Pi'(0)=-z^*(0)$ where $z^*(q)$ is the
value of $z$ for which $\mu(z)-qz$ is extremal. This gives
\begin{equation}
\beta_{eff}=2\frac{\gamma^{\delta}-1}{\gamma+\gamma^{\delta}}\frac{1}{T_-}
\quad \hbox{with} \quad
\gamma=\frac{T_+}{T_-}  ~; ~\delta=\frac{2}{2+D}.
\label{eq:beta}
\end{equation}
When $\gamma=1$
(i.e. if $\alpha=1$)
$\beta_{eff}=0$. As $\alpha$ decreases, $\gamma$ increases, since the
collisions dissipate more energy, and
$\beta_{eff} T_-$ grows, reaches a maximum and subsequently decreases
asymptotically toward $0$ as $\sim \gamma^{\delta-1}$.  We emphasize
that $\beta_{eff}$ does not depend upon $\omega$.  We have compared
these predictions with the numerical and experimental results,
measuring the temperatures $T_+$ and $T_-$ in the simulation. To
bypass the problem of temperature anisotropy (discussed below), we
have used the horizontal component of the temperature, obtaining (with
$\alpha=0.9$ and area fraction $13.8\%$) $T_+ \approx 141d^2f^2$ and
$T_- \approx 91d^2f^2$, i.e. $\gamma=1.55$ and, from Eq~(\ref{eq:beta})
in $D=2$, $\beta_{eff}=0.00193$ in very good agreement with the
measured value $0.0022$. It should be noted that the temperature ratio can
also be
approximated by $(\alpha^2)^{-m}$ where $m$ is the average number of
collisions undergone by a particle between the moments of entering and leaving
the observation region. In our numerical simulations (as well
as in the experiment) $m \approx 2$, which corresponds to $\gamma \approx
1.5$.

What happens for small values of $\tau$? We note that $\tilde{f}_Q(z)$ has the
form $\exp(\tau\mu(z))$ for {\em any} value of $\tau$ and not only for large
$\tau$. Therefore at small $\tau$ one can expand the exponential, obtaining
$\tilde{f}_Q(z) \sim 1+\omega\tau\left(-2+(1-T_+ z)^{-D/2}+(1+T_-
  z)^{-D/2}\right)$.  This immediately leads to an analytical expression for
$f_Q(Q_\tau)=\text{const} \times
\delta(Q_\tau)+\chi_{1,T_+}(Q_\tau)+\chi_{1,T_-}(-Q_\tau)$, which fairly
accounts for the strong peak which is observed in the experiment and in the
simulations, and predicts exponential tails for $f_Q(Q_\tau)$: $\chi_{1,T}(x)
\propto x^{D/2-1}\exp(-x/T)$ so that $\beta_+=1/T_+$ and $\beta_-=1/T_-$.
This suggests an experimental test of this theoretical approach: the measure
at small values of $\tau$ of the slopes of the exponential tails of
$f_Q(Q_\tau)$ should coincide with a direct measure of $T_+$ and $T_-$.
However, we point out that the values of $\beta_+$ and $\beta_-$ obtained by
fitting the tails in the hard disks simulation, using values as small as
$\tau=0.00015 \tau_{box}$ yield estimates of $T_+$ and $T_-$ which are larger
(by a factor $\sim 1.6$) than those found by a direct measure.  This
disagreement brings the limits of such a simple two-temperature picture to the
fore. In the simulation and in the original experiment the measured injected
energy is indeed the sum of several different contributions, namely $Q_\tau
\approx Q_\tau^{xx} + Q_\tau^{xy} + Q_\tau^{yx} + Q_\tau^{yy}$ where
$Q_\tau^{ij}$ is the kinetic energy transported by the $i$ component of the
velocity by particles crossing the boundary through a wall perpendicular to
direction $j$. Two main differences with the simplified interpretation given
above arise: a) there are {\em two} couples of temperatures, i.e.
$T_+^x,T_-^x$ as well as $T_+^y,T_-^y$~\cite{anisotropy}; b) the diagonal
contributions $Q_\tau^{jj}$ are sums of squares of velocities whose
distribution is not a Gaussian but is $\sim v \exp(-v^2/T)$, since the
probability of crossing is biased by the velocity itself. The calculation of
$f_Q(Q_\tau)$ is still feasible, with qualitatively similar
results~\cite{puglisi2}.

\begin{figure}[h]
\includegraphics[width=7cm,clip=true]{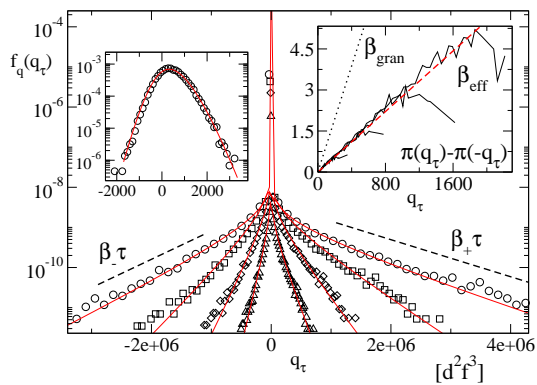}
\caption{
  {\bf Color online}: Pdf of transverse ($xy$) injected power $f_q(q_\tau)$
  for low values of $\tau=(1,2,4,8) \times \tau_{min}$ with
  $\tau_{min}=0.00015\;\tau_{box}$ (resp. circle, squares, diamond and
  triangles) from MD simulations with $N=270$ and $\alpha=0.9$. The
  distributions are shifted vertically for clarity. The (red) solid lines show
  the solution of the ``two temperatures'' model. The dashed lines
  indicate exponential tails $\sim \exp(\mp\beta_\pm\tau q_\tau)$. Left inset:
  same plot for a large value $\tau=6400\tau_{min}$. Right inset: plot of
  $(1/\tau)\log[f_q(Q_\tau)/f_q(-Q_\tau)]$ vs.  $q=Q/\tau$ from MD simulations
  (same parameters as above) at large values of $\tau=(1,2,4,8) \times
  \tau'_{min}$ with $\tau'_{min}=3200 \tau_{min}$, together with a dashed line
  of slope $\beta_{eff}$ predicted by Eq.~(\ref{eq:beta}) and the dotted line
  of slope $\beta_{gran}=1/T_{gran}$.
  \label{fig:pdf2}}
\end{figure}

Here we focus on the proposed two-temperature model,
showing its ability to explain the statistics of internal energy
currents. To this extent we have repeated the MD simulations
discussed above, but measuring only the transversal kinetic energy
current $Q_\tau^{xy}$ through the bottom boundary of the central
region. For new measurements, shown in Fig.~\ref{fig:pdf2}, we
have also improved the time resolution of the measure, using a minimum
$\tau=0.00015\tau_{cross}$.  The coefficients of the exponential tails
of $f_q(q_\tau)$ at low $\tau$ are in perfect agreement with the
predicted $\tau/T_\pm$ and the slope of the graph
$\log[f_q(q_\tau)/f_q(-q_\tau)]/\tau$ vs. $q_\tau$, shown in the
inset, is accurately recovered by Eq.~(\ref{eq:beta}) with $D=1$.  More
remarkably, the two-temperature  model (solved by numerical
inversion of $\tilde{f}$) fully accounts for the pdf $f_q(q_\tau)$
at {\em any} value of $\tau$, as evidenced in the same figure.

In conclusion we have implemented MD simulations of inelastic hard disks
fluidized in a vibrated box. We have studied the statistics of the power flux
through a closed perimeter --which effectively separates a high temperature
region from a low temperature one-- thereby accurately reproducing the
experimental measures of~\cite{feitosa}. Inspired by these results, we have put
forward a simple two-temperature Poissonian model, which entirely reproduces
the phenomenology of the system but for which the fluctuations of the energy
flow do not satisfy a Gallavotti-Cohen relation.
%in contrast with the initial interpretation of the experimental results~\cite{feitosa}. 
Moreover, our approach puts forward a kinetic interpretation for 
the measured slope $\beta_{eff}$, a question previously left open. We conclude here that
there are serious evidence against the power flux as a potential candidate for
extending the FR to a granular gas, a system for which the validity of such a
relation is therefore still open.

% A first order expansion of the large deviation function is sufficient to 
%obtain a quantitative agreement with numerical and experimental results. 
%An experimental test of the theoreticalapproach would consist in the 
%measure at small values of $\tau$, where experiments can gather 
%large statistics, of the exponential tails of $f_Q(Q_\tau)$: 
%the slopes should coincide with a direct measure of $T_+$ and $T_-$.

\begin{acknowledgments} We would like to thank B. Meerson
for useful discussions. A. P. acknowledges the Marie Curie grant No.
MEIF-CT-2003-500944. 
\end{acknowledgments}

\end{document}